\input{aipcheck}

\documentclass[
    ,final            % use final for the camera ready runs
  ]
  {aipproc}
  
\usepackage{epsf}
\usepackage{amssymb}
\usepackage{amsmath}
\usepackage{amsfonts}
\usepackage{psfrag,epsfig,graphicx}
\usepackage{graphicx}

\layoutstyle{8x11single}

\begin{document}

\title{Impact factor for high-energy two and three jets diffractive production}


\author{R.Boussarie}{
  address={LPT, Universit\'e Paris-Sud, CNRS, 91405, Orsay, France}
}

\author{A.V.Grabovsky}{
  address={Budker Institute of Nuclear Physics and Novosibirsk State University, 630090
Novosibirsk, Russia}
}

\author{L.Szymanowski}{
  address={National Centre for Nuclear Research (NCBJ), Warsaw, Poland}
}

\author{S.Wallon}{
  address={UPMC Univ. Paris 06, Facult\'e de Physique, 4 place Jussieu, 75252 Paris Cedex 05,
France
}
  ,altaddress={LPT, Universit\'e Paris-Sud, CNRS, 91405, Orsay, France} % additional visiting address
}

\begin{abstract}
 We present the calculation of the impact factor for the photon to quark, antiquark and gluon transition within Balitsky's shock-wave formalism. We also rederive the impact factor for photon to quark and antiquark transition. These results provide the necessary building blocks for further phenomenological studies of inclusive diffractive deep inelastic scattering as well as for two and three jets diffractive production which go beyond approximations discussed in the literature.
\end{abstract}

\maketitle

%%%%%%%%%%%%%%%%%%%%%%%%%%%%%%%%%%%%%%%%%%%%
%% MAINMATTER
%%%%%%%%%%%%%%%%%%%%%%%%%%%%%%%%%%%%%%%%%%%%

\section{Introduction}

One of the major achievements of HERA was the experimental evidence that 
among the whole set of $\gamma^* p \to X$ deep inelastic scattering events, almost 10\%  are diffractive (DDIS), of the form $\gamma^* p \to X Y$ with a rapidity gap between the proton remnants $Y$
and the hadrons $X$
coming from the fragmentation region of the initial virtual photon~\cite{Chekanov:2004hy-Chekanov:2005vv-Chekanov:2008fh,
Aktas:2006hx-Aktas:2006hy-Aaron:2010aa-Aaron:2012ad-Aaron:2012hua}.
Diffraction can be theoretically described according to several approaches, important for phenomenological applications. The first approach involves
a {\em resolved} Pomeron contribution (with a parton distribution function inside the Pomeron),  while the second one
relies on a  {\em direct} Pomeron contribution involving the coupling of a Pomeron with the diffractive state. The diffractive states can be modelled in perturbation theory by a  $q \bar{q}$ pair (for moderate $M^2$, where $M$ is the invariant mass of the  diffractively produced state $X$) or by higher Fock states as a $q \bar{q} g$ state for larger values of $M^2$. Based on such a model, with
 a two-gluon exchange picture for the Pomeron,  a good description of HERA data for diffraction could be achieved~\cite{Bartels:1998ea}. One of the important features of this approach is that the $q \bar{q}$ component with a longitudinally polarized photon plays a crucial role in the region of small
diffractive mass $M$, although it is a
twist-4 contribution.
In the direct components considered there, the $q \bar{q} g$ diffractive state has been studied in two particular limits. The first one, valid for very large $Q^2$, corresponds to a collinear approximation in which the transverse momentum of the gluon is assumed to be much smaller than the transverse momentum of the emitter~\cite{Wusthoff:1995hd-Wusthoff:1997fz}. 
The second one~\cite{Bartels:1999tn,Bartels:2002ri}, valid for very large $M^2$, is based on the assumption of a strong ordering of longitudinal momenta, encountered in BFKL equation~\cite{Fadin:1975cb-Kuraev:1976ge-Kuraev:1977fs-Balitsky:1978ic}. Both these approaches were combined in order to describe HERA data for DDIS~\cite{Marquet:2007nf}. 

Based on these very successful developments led at HERA
in order to understand the QCD dynamics with diffractive events, 
it would be appropriate to look for similar hard diffractive events at LHC. 
The idea there is to adapt the concept of  photoproduction of diffractive jets, which  was performed at HERA~\cite{Chekanov:2007rh,Aaron:2010su}, now with a flux of
quasi-real photons in ultraperipheral collisions (UPC)~\cite{Baltz:2007kq-Baur:2001jj}, relying on the notion of equivalent photon approximation. In both cases, 
 the hard scale is provided by the invariant mass of the tagged jets.

We here report on our computation~\cite{Boussarie:2014lxa} of the $\gamma^* \to q \bar{q} g$ impact factor at tree level with an arbitrary number of $t$-channel gluons described within the Wilson line formalism, also called QCD shockwave approach~\cite{Balitsky:1995ub-Balitsky:1998kc-Balitsky:1998ya-Balitsky:2001re}. As an aside, we rederive the $\gamma^* \to q \bar{q}$ impact factor. In particular, the 
$\gamma^* \to q \bar{q} g$ transition is computed without any soft or collinear approximation for the emitted gluon, in contrast with the above mentioned calculations. These results provide necessary generalization of building blocks for inclusive DDIS as well as for two- and three-jet diffractive production. Since the results we derived can account for an arbitrary number of $t$-channel gluons, this could allow to include higher twist effects which are suspected to be rather important in DDIS for $Q^2 \lesssim 5$ GeV$^2$~\cite{Motyka:2012ty}.

\section{Formalism}

As stated before, we use Balitsky's shockwave formalism. 
Its application shows that this method is very powerful in determining evolution equations and impact factors at next-to-leading order for inclusive processes~\cite{Balitsky:2010ze-Balitsky:2012bs}, at semi-inclusive level for $p_t$-broadening in $pA$ collisions~\cite{Chirilli:2011km-Chirilli:2012jd} or in the evaluation of the triple Pomeron vertex beyond the planar limit~\cite{Chirilli:2010mw}, when compared with usual methods based on summation of contributions of individual Feynman diagrams computed in momentum space. It is an effective way of estimating the effect of multigluon exchange. Its formulation in coordinate space makes it natural in view of describing saturation~\cite{GolecBiernat:1998js-GolecBiernat:1999qd}.
One introduces Wilson lines as 
\begin{equation}
U_{i}=U_{\vec{z}_{i}}=U\left(  \vec{z}_{i},\eta\right)  =P \exp\left[{ig\int_{-\infty
}^{+\infty}b_{\eta}^{-}(z_{i}^{+},\vec{z}_{i}) \, dz_{i}^{+}}\right]\,.
\label{WL}%
\end{equation}
The operator $b_{\eta}^{-}$ is the external shock-wave field built from slow gluons 
whose momenta are limited by the longitudinal cut-off defined by the rapidity $\eta$
\begin{equation}
b_{\eta}^{-}=\int\frac{d^{4}p}{\left(  2\pi\right)  ^{4}}e^{-ip \cdot z}b^{-}\left(
p\right)  \theta(e^{\eta}-|p^{+}|).\label{cutoff}%
\end{equation}
We use the light cone gauge
$\mathcal{A}\cdot n_{2}=0,$
with $\mathcal{A}$ being the sum of the external field $b$ and the quantum field
$A$%
\begin{equation}
\mathcal{A}^{\mu} = A^{\mu}+b^{\mu},\quad b^{\mu}\left(  z\right)  =b^{-}(z^{+},\vec{z}\,) \,n_{2}%
^{\mu}=\delta(z^{+})B\left(  \vec{z}\,\right)  n_{2}^{\mu}\,,\label{b}%
\end{equation}
where
$B(\vec{z})$ is a profile function.
%in transverse subspace.
The dipole operator 
$\mathbf{U}_{12}=\frac{1}{N_{c}}\rm{tr}\left(  U_{1}U_{2}^{\dagger}\right)  -1$
will be used extensively.

\section*{Impact factor for $\gamma\rightarrow q\bar{q}$ transition}

\begin{figure}
\scalebox{.89}{\begin{tabular}{cc}
\raisebox{3.75cm}{\includegraphics[scale=0.5]{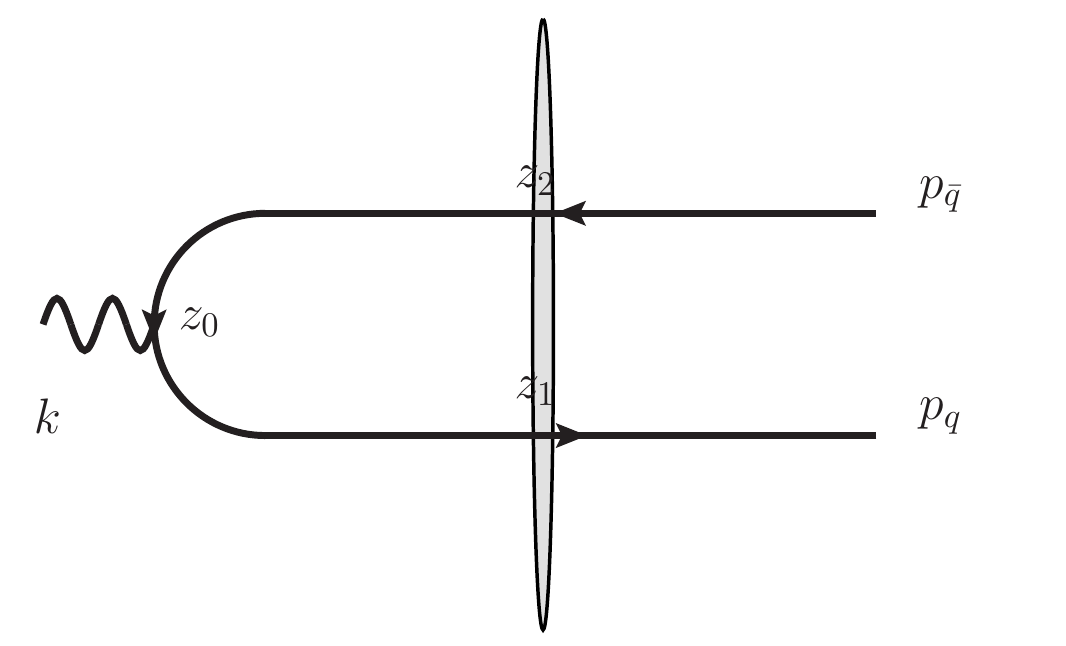}} &
\includegraphics[scale=0.65]{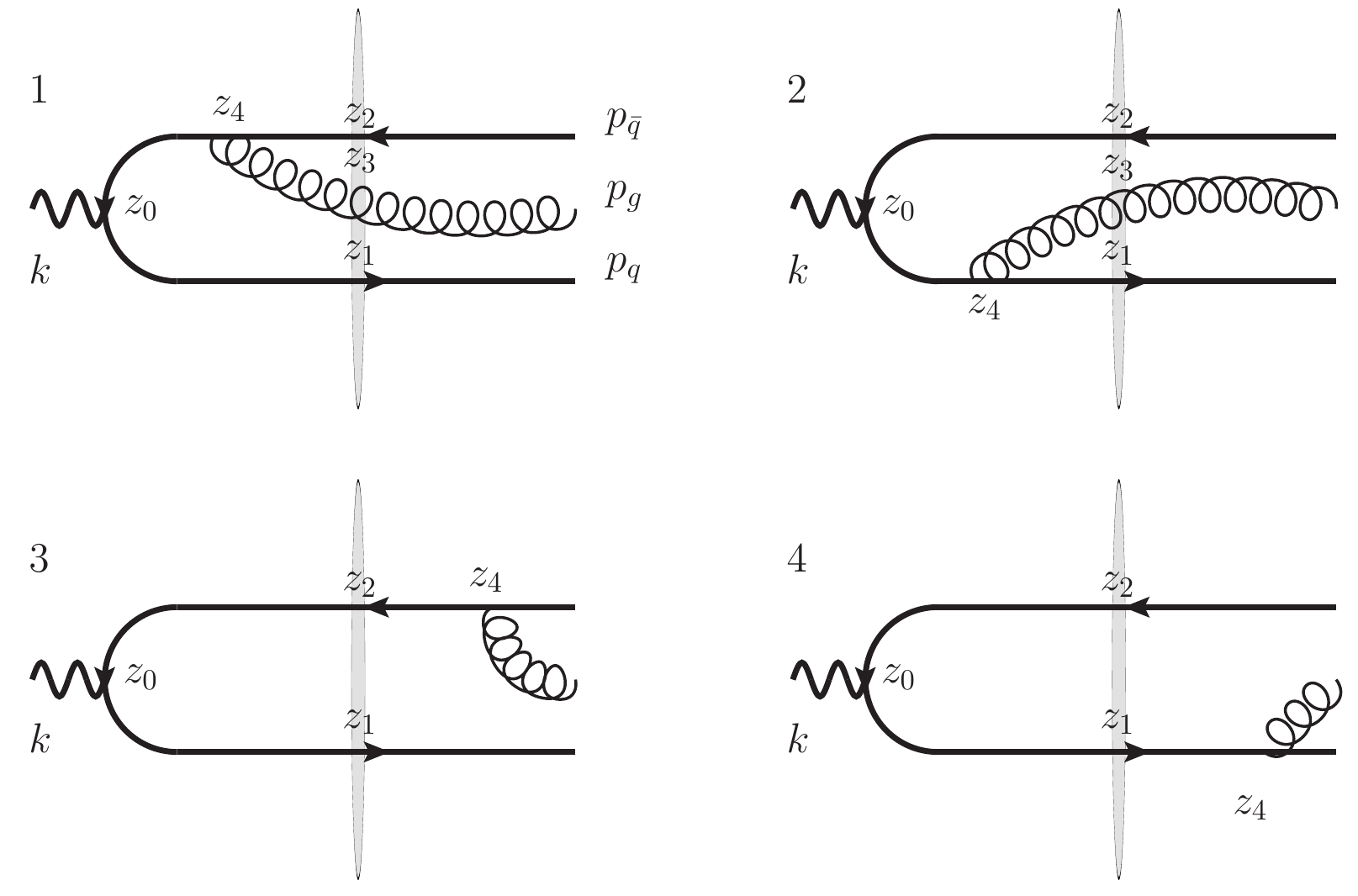}
\end{tabular}}
\caption{Left: diagram for $\gamma\rightarrow q\bar{q}$ transition. Right: the 4 diagrams for $\gamma\rightarrow q\bar{q}g$ transition.}
\label{Fig:diagrams}
\end{figure}

For $q \bar{q}$ production one can write, after projection on the color singlet state and subtraction of the non-interacting term
\begin{equation}
M_{0}^{\alpha}=N_c \int d\vec{z}_{1}d\vec{z}_{2}F\left(  p_{q},p_{\bar{q}}%
,z_{0},\vec{z}_{1},\vec{z}_{2}\right)  ^{\alpha} \mathbf{U}_{12}\,.
\label{M0int}%
\end{equation}
Denoting $Z_{12} = \sqrt{x_{q}x_{\bar{q}}\vec{z}_{12}^{\,\,2}}$, we get for a longitudinal photon
\begin{eqnarray}
\label{FL}
F\left(  p_{q},p_{\bar{q}},k,\vec{z}_{1},\vec{z}_{2}\right)  ^{\alpha
}\varepsilon_{L\alpha}=\theta(p_{q}^{+})\,\theta(p_{\bar{q}}^{+})\frac
{\delta\left(  k^{+}-p_{q}^{+}-p_{\bar{q}}^{+}\right)  }{(2\pi)^{2}}%
e^{-i\vec{p}_{q}\cdot \vec{z}_{1}-i\vec{p}_{_{\bar{q}}}\cdot\vec{z}_{2}}
(-2i)\delta_{\lambda_{q},-\lambda_{\bar{q}}}\,x_{q}x_{\bar{q}}%
\,Q\,K_{0}\left(Q \, Z_{12}\right)\,,
\end{eqnarray}
and for a transverse photon
\begin{eqnarray}
\label{FT}
F(  p_{q},p_{\bar{q}},k,\vec{z}_{1},\vec{z}_{2})  ^{j}%
\varepsilon_{Tj}\!=\theta(p_{q}^{+})\,\theta(p_{\bar{q}}^{+})\frac{\delta(
k^{+}\!\!-\!p_{q}^{+}\!-p_{\bar{q}}^{+}\!)  }{(2\pi)^{2}}e^{-i\vec{p}_{q}\cdot\vec
{z}_{1}-i\vec{p}_{_{\bar{q}}}\cdot\vec{z}_{2}}
\delta_{\lambda_{q},-\lambda_{\bar{q}}}( x_{q}-x_{\bar{q}%
}+s\lambda_{q})  \frac{\vec{z}_{12} \cdot \vec{\varepsilon}_{T}}{\vec{z}_{12}^{\,\,2}}
Q \,Z_{12} K_{1}(Q\, Z_{12})\,.\!\!\!\!\!
\end{eqnarray}

\section*{Impact factor for $\gamma\rightarrow q\bar{q}g$ transition}

For $q \bar{q} g$ production, projecting on the color singlet state and subtracting the non-interacting term again, one can write
\begin{eqnarray}
\nonumber 
M^{\alpha} &=& N_c^2 \int d\vec{z}_{1}d\vec{z}_{2}d\vec{z}_{3} \, F_{1}\left(  p_{q},p_{\bar{q}}%
,p_{g},z_{0},\vec{z}_{1},\vec{z}_{2},\vec{z}_{3}\right)  ^{\alpha}\frac{1}{2}
\left( \mathbf{U}_{32} + \mathbf{U}_{13} - \mathbf{U}_{12} + \mathbf{U}_{32}\mathbf{U}_{13} \right)
\\
&+& N_c \int d\vec{z}_{1}d\vec{z}_{2} \, F_{2}\left(  p_{q},p_{\bar{q}},p_{g},z_{0}%
,\vec{z}_{1},\vec{z}_{2}\right)  ^{\alpha}\frac{N_{c}^{2}-1}{2N_{c}} \mathbf{U}_{12}\,.
\label{F2tilde}%
\end{eqnarray}
The first and the second line of this equation correspond respectively to the two last diagrams of the first line and to the second line of diagrams of Fig.~\ref{Fig:diagrams}.
For a longitudinally polarized photon, they read
\begin{eqnarray}
&&\hspace{-1cm}F_{1}\left(  p_{q},p_{\bar{q}},p_{g},k,\vec{z}_{1},\vec{z}_{2},\vec{z}%
_{3}\right)  ^{\alpha}\varepsilon_{L\alpha}=2\,Q\,g\,\delta(k^{+}-p_{g}^{+}-p_{q}%
^{+}-p_{_{\bar{q}}}^{+})\theta(p_{g}^{+}-\sigma)\frac{e^{-i\vec{p}_{q} \cdot %
\vec{z}_{1}-i\vec{p}_{_{\bar{q}}} \cdot \vec{z}_{_{2}}-i\vec{p}_{g} \cdot \vec{z}_{3}}}%
{\pi\sqrt{2p_{g}^{+}}}
\nonumber \\
&\times&\delta_{\lambda_{q},-\lambda_{\bar{q}}}\left\{  (x_{_{\bar{q}}}%
+x_{g}\delta_{-s_{g}\lambda_{q}})x_{q}\frac{\vec{z}_{32} \cdot \vec{\varepsilon}%
_{g}^{\,\,\ast}}{\vec{z}_{32}^{\,\,2}}-(x_{q}+x_{g}\delta_{-s_{g}%
\lambda_{\bar{q}}})x_{_{\bar{q}}}\frac{\vec{z}_{31} \cdot \vec{\varepsilon}%
_{g}^{\,\,\ast}}{\vec{z}_{31}^{\,\,2}}\right\} K_{0}(QZ_{123}) \,,\\
\label{F1eL}%
\label{resF2L}
&&\hspace{-1cm}\tilde{F}_{2}\left(  p_{q},p_{\bar{q}},p_{g},k,\vec{z}_{1},\vec{z}_{2}\right)
^{\alpha}\varepsilon_{L\alpha}=4ig \, Q\,\theta(p_{g}^{+}-\sigma)\delta(k^{+}%
-p_{g}^{+}-p_{q}^{+}-p_{_{\bar{q}}}^{+})\frac{e^{-i\vec{p}_{q} \cdot \vec{z}%
_{1}-i\vec{p}_{_{\bar{q}}} \cdot \vec{z}_{2}}}{\sqrt{2p_{g}^{+}}}%
\nonumber \\
&&\times\delta_{\lambda_{q},-\lambda_{\bar{q}}}\frac{x_{q}\left(  x_{g}%
+x_{\bar{q}}\right)  \left(  \delta_{-s_{g}\lambda_{q}}x_{g}+x_{\bar{q}%
}\right)  }{x_{\bar{q}} \, x_g }\frac{\vec{P}_{\bar{q}} \cdot %
\vec{\varepsilon}_{g}^{\,\,\ast}}{\vec{P}_{\bar{q}}^2}\,e^{-i\vec{p}_{g} \cdot \vec{z}_{2}}K_{0}%
(QZ_{122})-\left(  q\leftrightarrow\bar{q}\right)  ,
\end{eqnarray}
while for a transversally polarized photon, we have 
\begin{eqnarray}
&&\hspace{-.7cm}F_{1}\left(  p_{q},p_{\bar{q}},p_{g},k,\vec{z}_{1},\vec{z}_{2},\vec{z}%
_{3}\right)  ^{\alpha}\!\varepsilon_{T\alpha}=\!-2i\,g\,Q\delta(k^{+}-p_{g}^{+}%
-p_{q}^{+}-p_{_{\bar{q}}}^{+})\theta(p_{g}^{+}-\sigma)
\frac{e^{-i\vec{p}_{q} \cdot \vec{z}_{1}-i\vec{p}_{_{\bar{q}}} \cdot \vec{z}_{_{2}%
}-i\vec{p}_{g} \cdot \vec{z}_{3}}}{\pi Z_{123}\sqrt{2p_{g}^{+}}}\delta_{\lambda
_{q},-\lambda_{\bar{q}}}K_{1}(QZ_{123})\\
&&\hspace{-.8cm}\times\left\{  \frac{\left(  \vec{z}%
_{23} \cdot \vec{\varepsilon}_{g}^{\,\,\ast}\right)  \left(  \vec{z}_{13} \cdot %
\vec{\varepsilon}_{T}\right)  }{\vec{z}_{23}{}^{2}}x_{q}\left(  x_{q}%
-\delta_{s\lambda_{\bar{q}}}\right)  \left(  x_{\bar{q}}+x_{g}\delta
_{-s_{g}\lambda_{q}}\right)  +\frac{\left(  \vec{z}_{23} \cdot \vec{\varepsilon}_{g}^{\,\,\ast}\right)
\left(  \vec{z}_{23} \cdot \vec{\varepsilon}_{T}\right)  }{\vec{z}_{23}{}^{2}}%
x_{q}x_{\bar{q}}\left(  x_{\bar{q}}+x_{g}\delta_{-s_{g}\lambda_{q}}%
-\delta_{s\lambda_{q}}\right)  \right\}  -\left(  q\leftrightarrow\bar
{q}\right) \, ,\nonumber\\
%\nonumber
\label{F1eT}%
\label{resF2tildeT}
&&\hspace{-.8cm}\tilde{F}_{2}\left(  p_{q},p_{\bar{q}},p_{g},k,\vec{z}_{1},\vec{z}_{2}\right)
^{\alpha}\varepsilon_{T\alpha}=-4g\,\theta(p_{g}^{+}-\sigma)\,\delta(k^{+}%
-p_{g}^{+}-p_{q}^{+}-p_{_{\bar{q}}}^{+})\frac{e^{-i\vec{p}_{q} \cdot \vec{z}%
_{1}-i\vec{p}_{_{\bar{q}}} \cdot \vec{z}_{2}}}{\sqrt{2p_{g}^{+}}}\delta_{\lambda
_{q},-\lambda_{\bar{q}}}%
\nonumber \\
&&\times  
\frac{\left(  \delta
_{\lambda_{\bar{q}}s}-x_{q}\right)  \left(  \delta_{-s_{g}\lambda_{q}}%
x_{g}+x_{\bar{q}}\right)}{x_{\bar{q}} \, x_g}  
\frac{\vec{P}_{\bar{q}} \cdot %
\vec{\varepsilon}_{g}^{\,\,\ast}}{\vec{P}_{\bar{q}}^2}
\frac{\vec{z}_{12} \cdot \vec{\varepsilon}_{T}}{\vec{z}_{12}^2} 
\, Q \, Z_{122}
K_{1}(QZ_{122})e^{-i\vec{p}_{g} \cdot \vec{z}_{2}%
}-\left(  q\leftrightarrow\bar{q}\right)  \,. 
\end{eqnarray}
We denote 
$ F_{2}\left(  p_{q},p_{\bar{q}},p_{g},z_{0},\vec{z}_{1},\vec{z}_{2}\right)^{\alpha}\!=\!\tilde{F}_{2}\left(  p_{q},p_{\bar{q}},p_{g},z_{0},\vec{z}_{1}%
 ,\vec{z}_{2}\right)  ^{\alpha}\!+\!\int d\vec{z}_{3}\,F_{1}\left(  p_{q},p_{\bar{q}%
 },p_{g},z_{0},\vec{z}_{1},\vec{z}_{2},\vec{z}_{3}\right)  ^{\alpha}\!.$

\section*{2- and 3-gluon approximation}

Let us notice that the dipole operator $\mathbf{U}_{ij}$ is of order $g^2$. Hence for only two or three exchanged gluons one can neglect the quadrupole term in the amplitude $M^{\alpha}$ and get 
\begin{eqnarray}
\label{M3gBis}
M^{\alpha} \overset{\mathrm{g^3}}{=}   \frac{1}{2}\int d\vec{z}_{1}d\vec{z}%
_{2} \mathbf{U}_{12}  \left[ \left(  N_{c}^{2}-1\right)
\tilde{F}_{2}\left(  \vec{z}_{1},\vec{z}%
_{2}\right) ^{\alpha} + \int d\vec{z}_{3} \left\{  N_{c}^{2}F_{1}\left(
\vec{z}_{1},\vec{z}_{3},\vec{z}_{2}\right)^{\alpha} +N_{c}^{2}F_{1}\left(  \vec{z}_{3},\vec{z}%
_{2},\vec{z}_{1}\right)  ^{\alpha} -  F_{1}\left(  \vec{z}_{1},\vec{z}_{2},\vec{z}_{3}\right)  ^{\alpha} \right\} \right].
\end{eqnarray}
For $\vec{p}_q=\vec{p}_g=\vec{p}_{\bar{q}}=\vec{0}$, those integrals can be performed analytically. Otherwise they can be expressed as a simple convergent integral over $[0,1]$ that can be performed numerically for any future phenomenological study. 

\section*{Conclusion}

The measurement of dijet production in DDIS was recently performed~\cite{Aaron:2011mp}, and a precise comparison of 
dijet versus triple-jet production, which has not been  performed yet at HERA~\cite{Adloff:2000qi}, would be very useful to get a deeper understanding of the QCD mechanism underlying diffraction. Recent investigations of the azimuthal distribution of dijets in diffractive photoproduction performed by ZEUS~\cite{Guzik:2014iba} show sign of a possible need for a 2-gluon exchange model, which is part of the shock-wave mechanism. Our calculation could be used for phenomenological studies of those experimental results.
A similar and very complementary study could be performed at LHC with UPC events. One should note that getting a full quantitative first principle analysis of this would  require an evaluation of virtual corrections to the $\gamma^* \rightarrow q\bar{q}$ impact factor, which are presently under study~\cite{Boussarie:prep}.

Diffractive open charm production was measured at HERA~\cite{Aktas:2006up} 
and studied in the large $M$ limit based on the direct coupling between a Pomeron and a $q \bar{q}$ or a $q\bar{q}g$ state, with massive quarks~\cite{Bartels:2002ri}. Such a program could also be performed at LHC, 
again based on UPCs and on
the extension of the above mentioned impact factors to the case of a massive quark. This could be further extended 
to $J/\Psi$  production, which are copiously produced at LHC.

\begin{theacknowledgments}
A. V. G.  acknowledges support of president grant MK-7551.2015.2 and
RFBR grant 13-02-01023. This work was partially supported by the PEPS-PTI PHENODIFF,
the PRC0731 DIFF-QCD, the Polish Grant NCN No. DEC-2011/01/B/ST2/03915, the ANR PARTONS (ANR-12-MONU-0008-01), the COPIN-IN2P3 Agreement
and the Joint Research Activity Study of Strongly Interacting Matter (acronym HadronPhysics3,
Grant Agreement n.283286) under the Seventh Framework Programme of the
European Community
\end{theacknowledgments}

\end{document}